\documentclass[12pt]{iopart}
\usepackage{graphics}
\usepackage{hyperref}
\usepackage{xfrac}
\usepackage{lscape,afterpage}
\usepackage{graphicx} 
\usepackage[font=small,labelfont=bf]{caption} 
\usepackage[subrefformat=parens]{subcaption}
\usepackage{amssymb}
\usepackage{multirow}
\usepackage[LGRgreek]{mathastext}
\begin{document}

\title[]{Excitation function measurement of $^{144}$Sm($\alpha$,n) reaction at sub-Coulomb energies and detailed covariance analysis}

\author{Tanmoy~Bar$^{1,2}$\footnote{Present affiliation: Inter-University Accelerator Centre, Aruna Asaf Ali Marg, New Delhi~-~110067, India}, Dipali~Basak$^{1,2}$, Lalit~Kumar~Sahoo$^{1,2}$, Sukhendu~Saha$^{1,2}$, Jagannath~Datta$^3$, Sandipan~Dasgupta$^3$, Chinmay~Basu$^{1,2}$}
\address{$^1$Nuclear Physics Division, Saha Institute of Nuclear Physics, 1/AF Bidhannagar, Saltlake City, Kolkata~-~700 064, India}
\address{$^2$Homi Bhabha National Institute, BARC Training School Complex, Anushaktinagar, Mumbai~-~400 094, India}
\address{$^3$Analytical Chemistry Division, Bhabha Atomic Research Centre; Variable Energy Cyclotron Centre, 1/AF Bidhannagar, Saltlake City, Kolkata~-~700 064, India}

\ead{amitanmoybar@gmail.com}

\vspace{10pt}
\begin{indented}
\item[]November 2025
\end{indented}

\begin{abstract}
The cross-section measurement of $^{144}$Sm($\alpha$,n)$^{147}$Gd (T$_{1/2}=$38.06(12) h) reaction has been performed at sub-Coulomb energies around 14$-$21 MeV ($V_{coul}\approx 21.8$ MeV) using the stacked foil activation technique. Irradiated targets were prepared from enriched (67\%) $^{144}$Sm$_2$O$_3$ powder using molecular deposition technique between thickness 280$-$350 $\mu$g/cm$^2$ on high purity Al backing. A detailed simulation has been carried out to address the energy uncertainty in the irradiated beam energy followed by a comprehensive discussion of various uncertainties in the form of covariance and correlation matrices. Finally the excitation functions are compared with the previously measured experimental data from literature and the theoretical predictions obtained using Hauser-Feshbach statistical model code.
\end{abstract}

%
%Uncomment for keywords
\vspace{2pc}
%
% Uncomment for Submitted to journal title message
\submitto{\JPG}
%
% Uncomment if a separate title page is required
%\maketitle
% 
% For two-column output uncomment the next line and choose [10pt] rather than [12pt] in the \documentclass declaration
%\ioptwocol
%

\section{Introduction}
Cross section of nuclear reactions induced by light charged particles (e.g. proton, alpha) are of great significance in many fields like nuclear physics, astrophysics, medical radioisotope studies etc~\cite{qaim2016uses,uddin2011excitation, choudhary2022measurement, bar2024measurement, basak2024experimental, saha2025proton, upadhyay2025production, hermanne2018reference, mukhopadhyay2011applications}. The $^{144}$Sm($\alpha$,n)$^{147}$Gd reaction is of considerable significance in nuclear astrophysics, particularly in studies of the p-nuclei. Moreover, the produced $^{147}$Gd isotope is of interest in single-photon emission tomography (SPET) as a potential magnetopharmaceutical~\cite{denzler1997production,Denzler1995}. Of all other radioisotopes of gadolinium, $^{147}$Gd is the most favourable isotope for SPET imaging, because of its suitable half-life of 38.06 hours and a strong $\gamma-$ray of 229.32 keV (60.7\%)~\cite{Nica2022}. This ($\alpha$,n) reaction is one of the major channel to produce $^{147}$Gd.
In the context of nuclear astrophysics, there are around 35 neutron deficit nuclei exist, beyond Fe, between $^{74}$Se and $^{196}$Hg which are called \textit{p-}nuclei~\cite{arnould2003p,woosley1978p}. They does not produce via \textit{s} or \textit{r-} process and have typically much lower abundances in the solar system. The typical abundance is between 0.01$-$0.3\% for nuclei having atomic number, Z$>$50. But the neutron magic \textit{p-}nuclei $^{144}$Sm and $^{92}$Mo are much more abundant, having 3.08\% and 14.52\% respectively. The abundance of any nucleus is governed by a network of reactions contributing to its production and destruction pathways. Consequently, investigating the reactions involving $^{144}$Sm in astrophysical environments is essential for accurate abundance calculations. $^{144}$Sm($\alpha$,n)$^{147}$Gd cross section measurement at below Coulomb barrier energies is a useful contribution towards the astrophysical relevance modelings as well as investigation of medically important $^{147}$Gd radioisotope.
The cross section values determined using the method of activation, can have uncertainties from various sources. It is extremely important to account all these contributions carefully. The correlation (covariance) among the measurements has to also be considered in order to avoid overestimation or underestimation of the uncertainty in the quantity of interest (eg. reaction rate). In the present study, a detailed discussion on covariance matrix calculation has been done for the first time for this reaction. The cross correlation between measured cross sections are calculated by considering the uncertainties in incident beam flux, $\gamma-$ray intensity, target thickness etc. The measured excitation functions were compared with the previously available, though limited, data sets~\cite{gyurky2023cross, Denzler1995, archenti1989alpha} and with theoretical predictions calculated using all parameter combinations implemented in the TALYS-2.0 reaction code~\cite{koning2023talys}. Another important reaction channel at these energies is $^{144}Sm(\alpha,\gamma)^{148}Gd$. However, in the present activation approach, it cannot be identified via offline $\gamma-$ray spectroscopy, as $^{148}Gd$ decays exclusively by $\alpha$ emission ($T_{1/2}\approx$ 75 years)~\cite{Nica2026}.

\section{Experiment}

The experiment was performed at VECC, Kolkata using the stacked foil activation technique. It was then followed by an offline $\gamma-$ray spectroscopy measurement. Two stacks were individually bombarded with 28 MeV alpha beam. After degradation, the cross section measurements were carried out for effective alpha energies between 14 and 21 MeV. Prior to entering the target stack, the alpha-beam energy was reduced using aluminium degrader foils. $^{144}$Sm targets (total of five) were irradiated for 17-22 hours with a average $^4He^{2+}$ beam current of $\sim$210$-$270 nA. Details of the experimental procedure and data analysis are described below.
\subsection{Target preparation}

The targets were prepared using molecular deposition technique~\cite{parker1964molecular,bar2022preparation} from 67\% isotopically enriched $^{144}$Sm$_2$O$_3$ powder. 10 mg of enriched samariam oxide power was utilised to prepared approximately 6-8 targets using a in-house made target deposition setup. High purity Al foil (99.45\%) of 25 $\mu$m thickness was used as a backing. The target preparation procedure is described in Refs.~\cite{bar2022preparation,bar2024measurement}. Targets of thicknesses between 280-350 $\mu$g/cm$^2$ were prepared for this cross section measurement. The initial thickness of the prepared foils was estimated by weighing the backing foils before and after deposition. Subsequently, a $^{229}$Th alpha source (8.376 MeV) was used to determine the target thickness, yielding an uncertainty in the range of 15–25\%.

\subsection{Preparation of the target stacks and irradiation setup}
\begin{figure}
  \centering
      \includegraphics[clip, trim=0.0cm 0.0cm 0.0cm 0.0cm,width=0.6\linewidth]{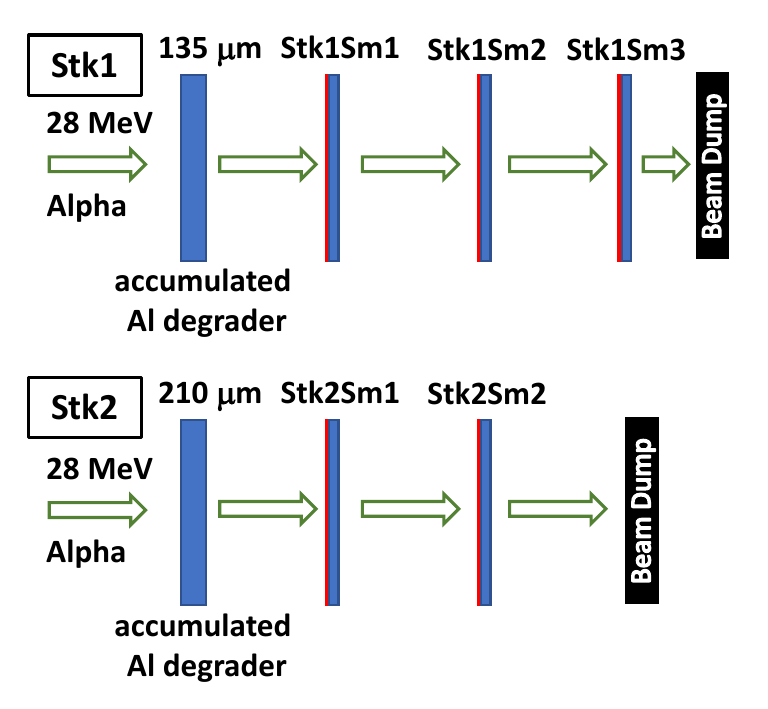}
  \caption{Target stack setup and target positions for multiple target irradiation.}
  \label{stack_setup}
\end{figure}

\begin{figure}
  \centering
    \includegraphics[clip, trim=0.0cm 0.0cm 0.0cm 0.0cm,width=0.85\linewidth]{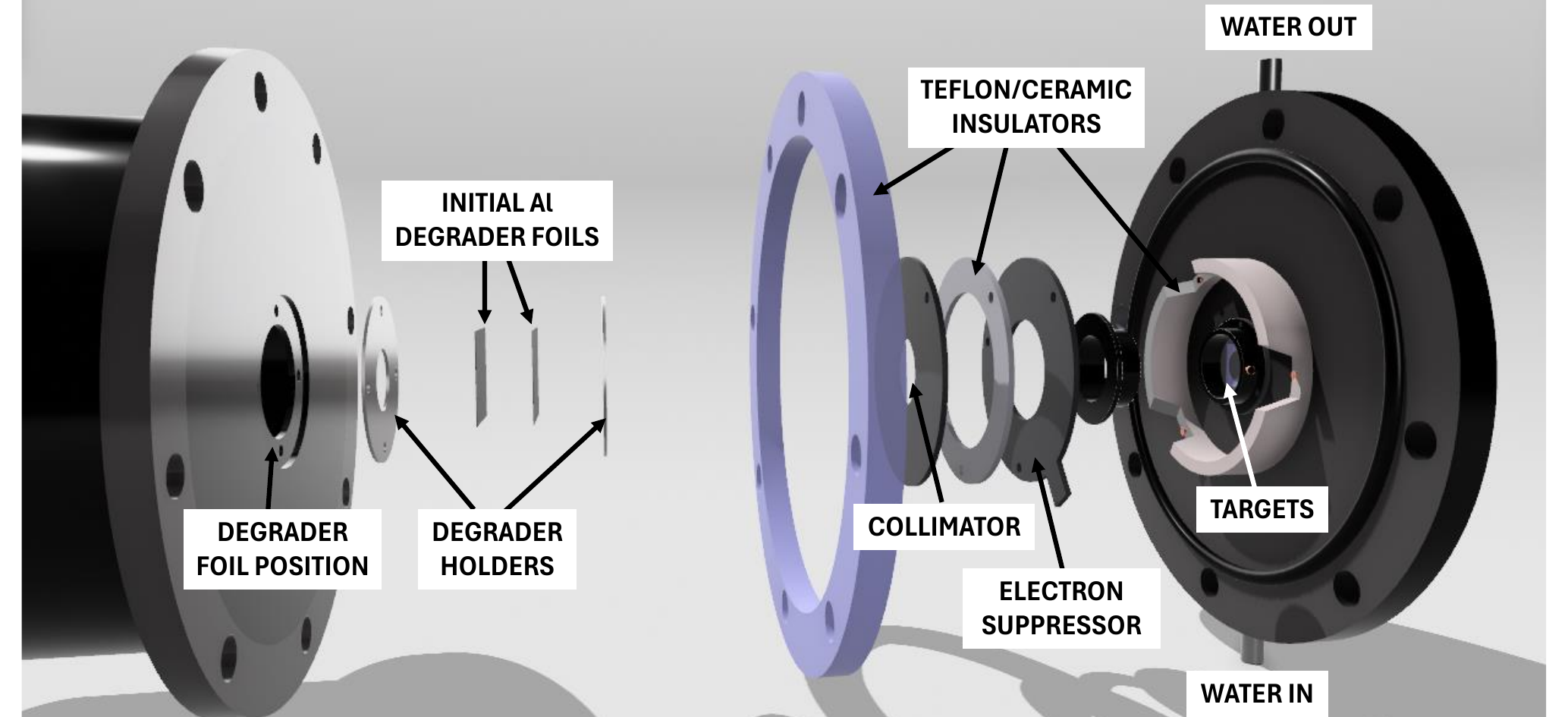}
    \caption{Targets irradiation setup with electron suppressor and water cooled flange.} 
    \label{setup3} 
  \end{figure}

$^{144}$Sm($\alpha$,n) reaction cross section for five energies below the Coulomb barrier ($\sim$21.8 MeV) have been measured using two stacks of targets between 14$-$21 MeV (Figure~\ref{stack_setup}). First stack (Stk1) targets were irradiated for energies between 21 and 18 MeV (three energy data points) and second stack (Stk2) for 15$-$14 MeV (two energy data points). Both the stacks consisted of initial Al foils as energy degrader having thicknesses 135 $\mu$m and 210 $\mu$m respectively. During irradiation, the target holder flange was cooled by circulating low-conductivity water. The beam current values were logged every minute with a current integrator to monitor variations during bombardment. In the experimental setup, an additional plate biased at $-500$ V was positioned in front of the target stack to serve as an electron suppressor. The complete irradiation arrangement used in this measurement is shown in Figure~\ref{setup3}.

\subsection{Irradiation energy determination using \texttt{GEANT4}}

The lowest alpha-beam energy available from the cyclotron was 28 MeV with an energy spread of about 200 keV (FWHM). The beam was initially passed through aluminium degrader foils to reduce its energy, and after each $^{144}$Sm$_2$O$_3$ target, a 25-µm pure aluminium backing foil served as an additional degrader for the subsequent targets. Such multistage energy degradation introduces uncertainties in the effective irradiation energy at each foil. To quantify these effects, a detailed Monte Carlo simulation was performed using the \texttt{GEANT4} framework~\cite{apostolakis2007overview}, accounting for all relevant energy straggling processes, including the initial beam-energy spread from the cyclotron.

A detailed geometry of the foil arrangement for both stacks (Stk1 and Stk2) was implemented within the GEANT4 simulation framework (\texttt{G4Construction}). A total of $10^5$ alpha particles with an initial energy of 28 MeV and a 0.2 MeV FWHM energy spread (generated using a uniform random distribution) were projected randomly over an 8 mm diameter circular area to reproduce the experimental beam spot.

The interactions of alpha particles with the Sm$_2$O$_3$ target layer, together with the aluminium backing and initial degrader foils, were modelled using the physics lists \texttt{G4HadronElasticPhysics}, \texttt{G4HadronPhysicsQGSP\_BIC} (for light-ion reactions), \texttt{G4IonPhysics} (for light projectiles such as protons and alpha particles), and \texttt{G4StoppingPhysics} for energy-loss calculations. The number of simulated alpha particles was progressively increased from $10^2$ to $10^5$ to reduce statistical uncertainties and to obtain well-defined Gaussian fits for extracting the mean energy and energy straggling (1$\sigma$) at each target position.

\begin{figure}
  \centering
    \includegraphics[clip, trim=0.0cm 0.0cm 0.0cm 0.0cm,width=0.8\linewidth]{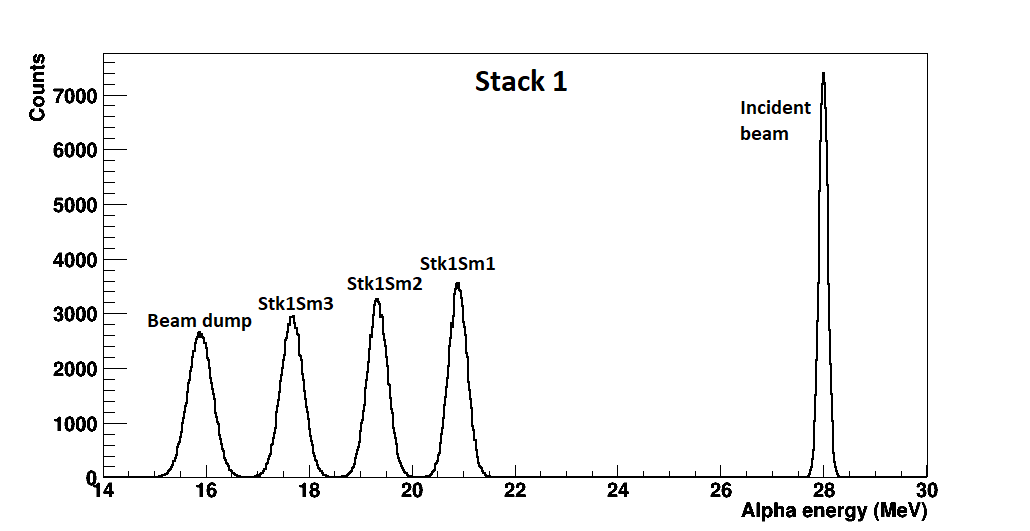}
    \caption{Effective irradiation energy and energy straggling for the Stack 1 targets obtained from GEANT4 simulations.} 
    \label{geant4-1} 
  \end{figure}

\begin{figure}
  \centering
    \includegraphics[clip, trim=0.0cm 0.0cm 0.0cm 0.0cm,width=0.8\linewidth]{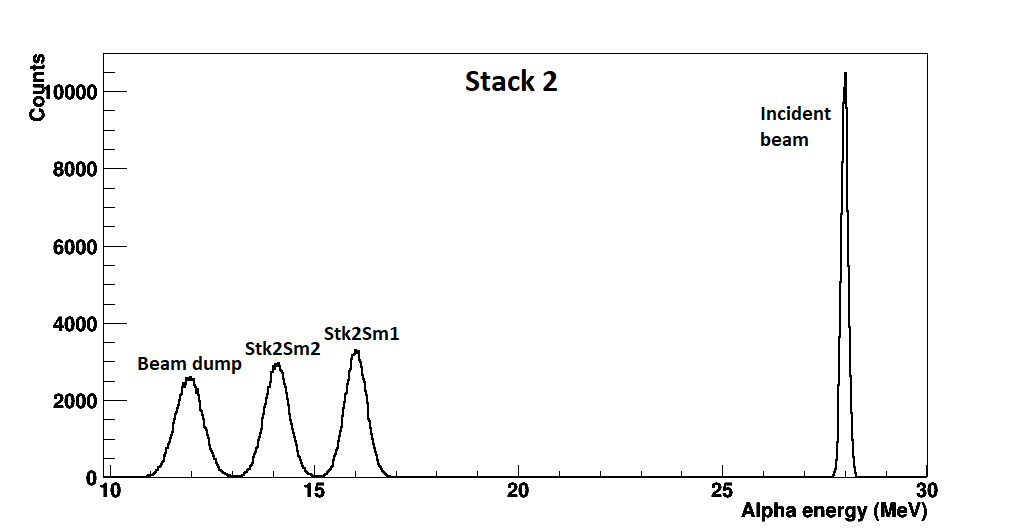}
    \caption{Same as Figure~\ref{geant4-1} but for Stack 2 targets.} 
    \label{geant4-2} 
  \end{figure}

The resulting energy distributions were analysed using the CERN ROOT package~\cite{brun1997root}. Figures~\ref{geant4-1} and~\ref{geant4-2} present the simulated results for Stk1 and Stk2, respectively.

The simulation results (Figure~\ref{geant4-1} \&~\ref{geant4-2}) indicate that as the beam traverses successive target foils, it undergoes progressively greater energy loss. This cumulative degradation also contributes to an increased uncertainty in the bombardment energy. The energy corresponding to the 1$\sigma$ width of the distribution was taken as the measure of energy straggling in the incident beam for each target.

\subsection{Measurement of $\gamma-$ray activity}

\begin{figure}
  \centering
    \includegraphics[clip, trim=0.0cm 0.0cm 0.0cm 0.0cm,width=0.8\linewidth]{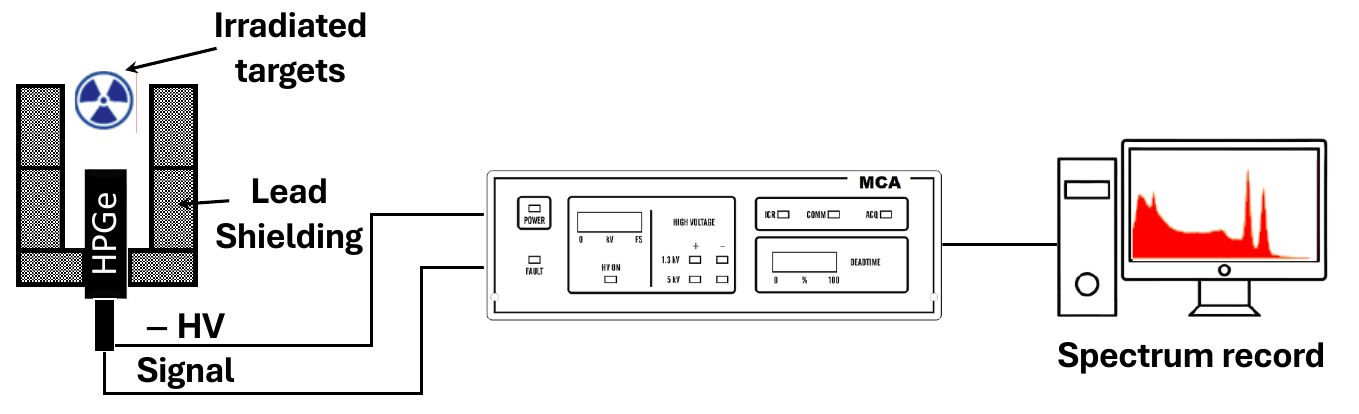}
    \caption{Schematic diagram of detection setup used for $\gamma-$ spectroscopy measurement.} 
    \label{detsetup} 
  \end{figure}

After irradiation, the stacks (Stk1 and Stk2) were allowed to cool for several hours to reduce the high residual activity arising from reactions in Sm, Al, and surrounding materials, thereby minimizing radiation hazards. This cooling time before the measurement also reduces the unwanted background activity due to the backing Al foil. The $\gamma-$activity measurement was done using a high-purity germanium (HPGe) detector having $\sim$1.8 keV energy resolution at 1.33 MeV $\gamma-$energy and 40\% relative efficiency. A 7.5 cm thick lead brick wall was used to reduce the room background during the counting. The data acquisition was carried out using a full featured 16K channel integrated multichannel analyzer (MCA) based on digital signal processing technology, DSA (Digital Signal Analyzer) and a spectroscopy software suite for spectrum recording in the computer. Figure~\ref{detsetup} shows a schematic setup diagram of the complete detection system. 

\subsection{Detector efficiency calibration}

HPGe detector used in this off-line gamma activity measurement was calibrated with a standard $^{152}$Eu point-like source having half-life, T$_{1/2}$=13.517$\pm$0.009 years~\cite{MARTIN20131497}. The absolute photo peak efficiency of the detector was assessed using the known activity of the $^{152}$Eu (A${_0}$=3.908 $\times$10$^4$ Bq) on 17 May 1982.

The efficiency of the detector for the point source ($\epsilon_{point}$) was determined using~\cite{PhysRevC.95.024619}

\begin{equation}
\epsilon_{p}=\frac{KC}{I_\gamma~A_0~e^{-\lambda t}~t_m}
\label{eff_eq}
\end{equation}
where $C$ is the count under the $\gamma$ peak, K is the correction factor to counter any summing effect, A$_0$ is the $^{152}$Eu source activity at the time of production, t is the time elapsed between Eu-source production (when activity was A$_0$) and start of counting time. I$_\gamma$ and $\lambda$ ($=\frac{0.693}{T_{1/2}}$, $T_{1/2}$ being half-life of nuclei) represents decay intensity and decay constant respectively and t$_m$ is the counting time.

First two irradiated targets from stack1 (Stk1Sm1 and Stk1Sm2) were measured at 50 mm distant from the detector surface, rest of the foils were measured at close geometry configuration (12.5 mm) to maximise the statistics. The efficiency of the detector was measured for both the distances. Such close geometries can give rise to the possibility of co-incidence summing specially for the measurements taken at 12.5 mm. To improve the accuracy of the efficiency determination, summing-effect corrections and extended-geometry corrections for the target samples were implemented, as discussed in the next section.

\subsection{Efficiency correction due to summing and sample geometry}

The co-incidence summing occurs when two $\gamma-$rays emitted in cascade enter the detector active medium within the resolving time of the detector. In such scenario, the detector records them as a single event with an energy equal to the sum of the two photon energies. Such events can mislead the count of $\gamma-$ rays of interest. For any source and detector configuration,  some degree of coincidence summing is always present. This effect becomes more pronounced when the source is positioned close to the detector surface. The summing effect is incorporated by introducing a correction factor (K) in the efficiency calculation (Eq.~\ref{eff_eq}). This factor was evaluated using the Monte Carlo code EFFTRAN (version 4.7.1), incorporating detailed specifications of the source and detector geometries. The higher K values observed for the 12.5 mm configuration indicate stronger summing effects compared to the 50 mm geometry.

Along with summing factor, another correction in the efficiency values are done due to the finite geometry of the irradiated targets unlike the standard source which is point like utilized for efficiency calibration. The irradiated samples exhibited an active area of $\sim$8 mm diameter, defined by the beam spot. The efficiency values obtained using point-source standards, after applying summing corrections, were subsequently employed to determine the detection efficiency for finite-dimension samples using the EFFTRAN code.

\begin{table}
\centering
\caption{\label{tab_eff12.5mm}Efficiency correction for summing effect ($K$), point source efficiency($\varepsilon_p$), and corrected efficiency for the extended sample ($\varepsilon$) at a source–detector distance of 12.5 mm.}
\footnotesize
\begin{tabular}{@{}cccccc}
\br
$\gamma-$energy (Intensity\%)&Area under peak ($C$)&$K$&$\varepsilon_p$& $\varepsilon$\\
\mr
121.8 (28.53$\pm$0.16) 	&467836(0.19\%)	&1.131	&0.1293	&0.1254(0.93\%)\\
\mr
244.7 (7.55$\pm$0.04) 	&79367(0.48\%)	&1.193	&0.0874	&0.0847(1.00\%)\\
\mr
344.3 (26.59$\pm$0.02) 	&262917(0.25\%)	&1.088	&0.0750	&0.0727(0.77\%)\\
\mr
411.1 (2.237$\pm$0.013)	&15032(1.07\%)	&1.223	&0.0573	&0.0556(1.39\%)\\
\mr
444.0 (3.125$\pm$0.018) &21120(0.84\%)	&1.170	&0.0551	&0.0534(1.22\%)\\
\mr
778.9 (12.93$\pm$0.08) 	&63148(0.51\%)	&1.129	&0.0384	&0.0372(1.07\%)\\
\mr
964.1 (14.51$\pm$0.07)	&60747(0.52\%)	&1.092	&0.0319	&0.0309(1.00\%)\\
\mr
1112.1 (13.67$\pm$0.08)	&53163(0.55\%)	&1.049	&0.0284	&0.0276(1.07\%)\\
\mr
1408.0 (20.87$\pm$0.09)	&65180(0.51\%)	&1.068	&0.0233	&0.0226(1.00\%)\\
\br
\end{tabular}
\end{table}
\begin{table}
\centering
\caption{\label{tab_eff50mm}Same as Table~\ref{tab_eff12.5mm} at a source–detector distance of 50 mm.}
\footnotesize
\begin{tabular}{@{}cccccc}
\br
$\gamma-$energy (Intensity\%)&Area under peak ($C$)&$K$&$\varepsilon_p$& $\varepsilon$\\
\mr
121.8 (28.53$\pm$0.16) 	&368705(0.16\%)	&1.034	&2.694	&2.786(0.56\%)\\
\mr
244.7 (7.55$\pm$0.04) 	&66549(0.39\%)	&1.048	&1.838	&1.926(0.75\%)\\
\mr
344.3 (26.59$\pm$0.02) 	&184538(0.23\%)	&1.023	&1.447	&1.480(0.61\%)\\
\mr
411.1 (2.237$\pm$0.013)	&13254(0.87\%)	&1.054	&1.235	&1.302(1.33\%)\\
\mr
444.0 (3.125$\pm$0.018) &17094(0.76\%)	&1.043	&1.140	&1.189(1.20\%)\\
\mr
778.9 (12.93$\pm$0.08) 	&46615(0.46\%)	&1.032	&0.752	&0.776(0.83\%)\\
\mr
964.1 (14.51$\pm$0.07)	&44819(0.47\%)	&1.024	&0.644	&0.659(0.84\%)\\
\mr
1112.1 (13.67$\pm$0.08)	&36714(0.52\%)	&1.013	&0.560	&0.567(0.90\%)\\
\mr
1408.0 (20.87$\pm$0.09)	&46230(0.47\%)	&1.018	&0.462	&0.462(0.84\%)\\
\br
\end{tabular}
\end{table}

\begin{figure}[h]
\centering
\includegraphics[clip, trim=0.0cm 0.0cm 0.0cm 0.0cm,width=0.6\textwidth]{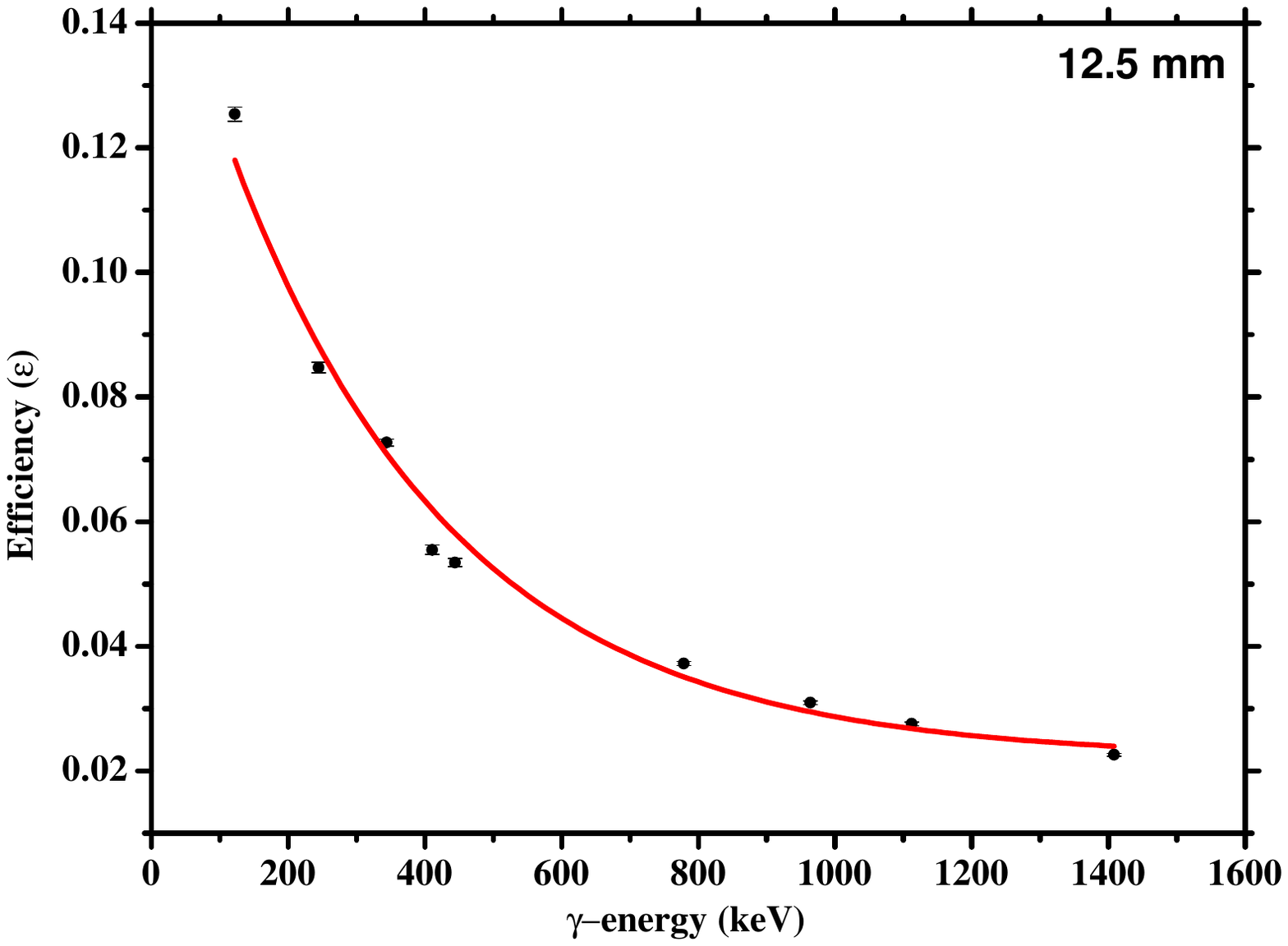}
\caption{Efficiency curve for 12.5 mm distance.}
\label{eff_curve12.5}
\end{figure}

\begin{figure}[h]
\centering
\includegraphics[clip, trim=0.0cm 0.0cm 0.0cm 0.0cm,width=0.6\textwidth]{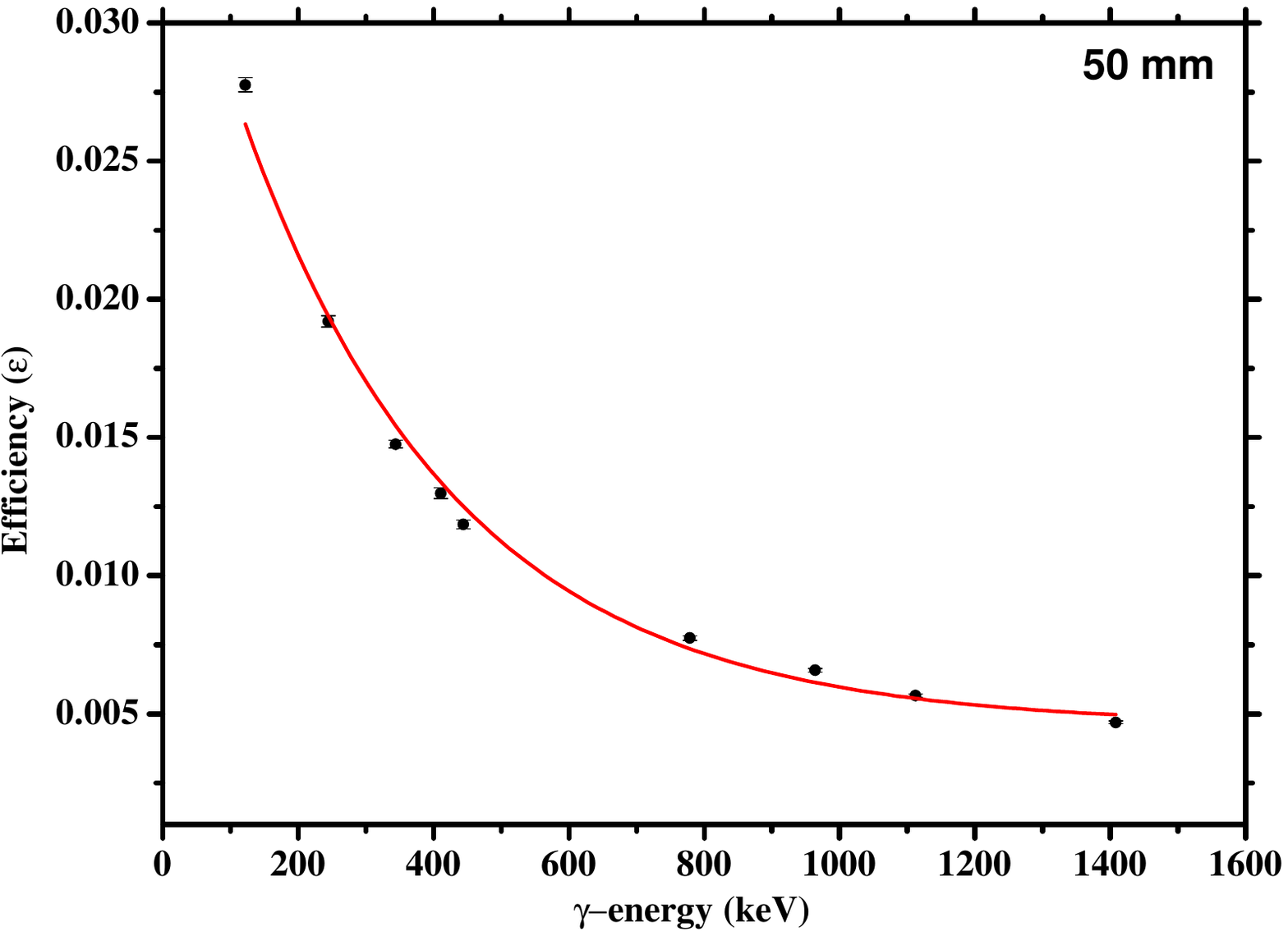}
\caption{Efficiency curve for 50 mm distance.}
\label{eff_curve50}
\end{figure}

\subsection{Detector efficiency curve}

After incorporating corrections due to summing effect and finite dimension correction to efficiency, the relation between the final efficiency ($\epsilon$) and $\gamma-$energy is derived for two positions by fitting the data shown in Table~\ref{tab_eff12.5mm} and Table~\ref{tab_eff50mm}. The data is fitted with the functional form of efficiency mentioned in Eq.~\ref{eff_fit}.\\
The fitting parameters and corresponding covariance matrix is mentioned in Table~\ref{eff_fitTab12.5mm} and Table~\ref{eff_fitTab50mm}. Detailed of covariance matrix is discussed in later sections.

\begin{equation}
\varepsilon (E_\gamma)=\varepsilon_1 e^{-E_\gamma/E_0}+\varepsilon_0
\label{eff_fit}
\end{equation}
The fitting parameters ($\varepsilon_1$, E$_0$ and $\varepsilon_0$) are obtained from the detector efficiencies measured using $^{152}$Eu standard source.
\begin{table}
\centering
\caption{\label{eff_fitTab12.5mm}Efficiency curve fitting parameters for 12.5 mm distance}
\footnotesize
\begin{tabular}{@{}cclll}
\br
Parameter&Fitted value&&Covariance matrix&\\
\mr
$\varepsilon_0$&0.002205&3.78848$\times$10$^{-6}$&&\\
\mr
$\varepsilon_1$&0.13897&1.24724$\times$10$^{-5}$&1.62665$\times$10$^{-4}$\\
\mr
$E_0$&329.427 &$-$0.06198&$-$0.41217&1502.97643\\
\br
\end{tabular}
\end{table}
\begin{table}
\centering
\caption{\label{eff_fitTab50mm}Efficiency curve fitting parameters for 50 mm distance}
\footnotesize
\begin{tabular}{@{}cclll}
\br
Parameter&Fitted value&&Covariance matrix&\\
\mr
$\varepsilon_0$&0.00458&1.00589$\times$10$^{-7}$&&\\
\mr
$\varepsilon_1$&0.03187&3.55521$\times$10$^{-7}$&5.16874$\times$10$^{-6}$\\
\mr
$E_0$&319.27841 &$-$0.00737&$-$0.05412&824.0464\\
\br
\end{tabular}
\end{table}
This relations were used to determine detector efficiency at specific gamma energy coming from the activated sample targets at specific distance from detector surface.

\section{Data Analysis}
\subsection{Calculation of reaction cross section}
\begin{figure}
  \centering
    \includegraphics[clip, trim=0.0cm 0.0cm 0.0cm 0.0cm,width=0.8\linewidth]{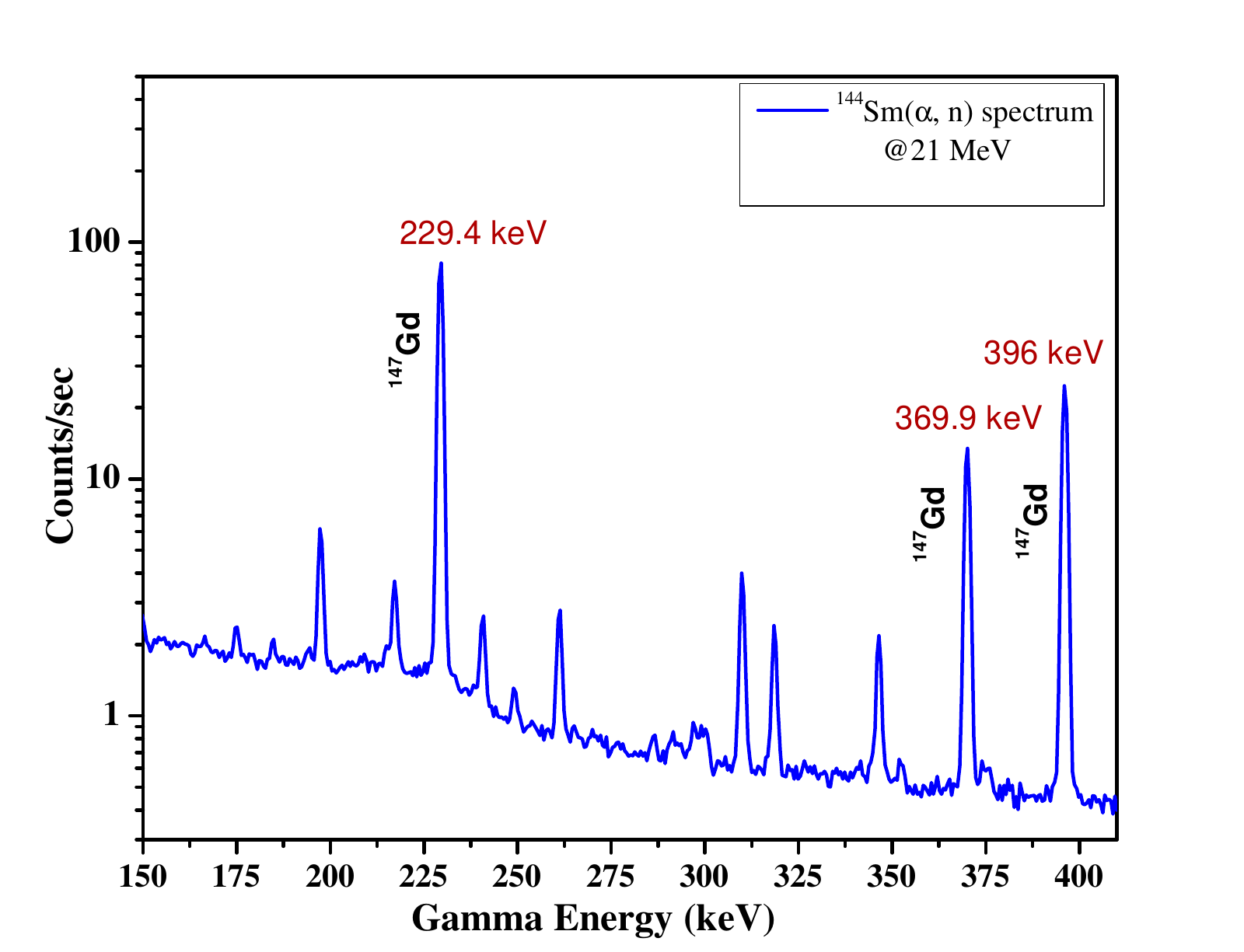}
    \caption{Typical gamma spectroscopy from irradiated $^{144}$Sm target at $\sim$21 MeV. Prominent $\gamma-$rays related to the nuclei of interest are marked.} 
    \label{spec} 
  \end{figure}

The neutron-emission channel cross section of $^{144}$Sm$+\alpha$ (S$_n=$ 12.6 MeV) reaction was measured for five different incident energies. $\gamma-$activity measurement of irradiated targets was performed for 300 secs to $5\frac{1}{2}$ hours to obtain significant statistics. A typical normalised spectrum is shown in Figure~\ref{spec} with $\gamma-$rays related to $^{147}$Gd marked. Most intense 229.3 keV $\gamma-$transition having absolute intensity 60.7$\pm$3.0\%~\cite{Nica2022} has been considered for cross section calculation using the following relation, 
\begin{equation}
\sigma = \frac{C~\lambda}{\phi_b~N_{target}~I_\gamma~\varepsilon_{det}~(1-e^{-\lambda t_{irr}}) (e^{-\lambda t_{cool}} - e^{-\lambda(t_{cool} + t_m)})}
\label{cross_eq3}
\end{equation}
where $C$ is the count under the peak, $\varepsilon_{det}$ stands for detector efficiency, $t_{cool}$ is the time between end of irradiation and start of counting, and $t_m$ is the counting time. 
where $\phi_b$ denotes proton flux (1/s), $\sigma$ is the cross-section (cm$^2$), $N_{target}$ is the target atoms/cm$^2$ and $t_{irr}$ is the irradiation time.

All irradiated energies and corresponding energy straggling (1$\sigma$) due to energy degradation by foils placed upfront have been listed in Table~\ref{irr_energy}.
\begin{table}
\centering
\caption{\label{irr_energy}The irradiated energies of $^{144}$Sm for both stacks and the corresponding energy uncertainties (1$\sigma$), as calculated using GEANT4, are presented. The accelerator beam energy uncertainty of 200 keV (FWHM) has also been included, see Figure~\ref{stack_setup} for stack setup.}
\footnotesize
\begin{tabular}{@{}cccc}
\br
&Irradiated energy&Irradiated Target&Energy uncertainty (1$\sigma$)\\&(MeV)&(see Figure~\ref{stack_setup})& (MeV)\\
\mr
Stack~1	&20.90	&Stk1Sm1	&$\pm$ 0.18\\
		&19.34	&Stk1Sm2	&$\pm$ 0.20 \\
		&17.68	&Stk1Sm3	&$\pm$ 0.22 \\
\mr
Stack~2	&16.03	&Stk2Sm1	&$\pm$ 0.27\\
		&14.09	&Stk2Sm2	&$\pm$ 0.30 \\

\br
\end{tabular}
\end{table}

%%\begin{table}
%%\centering
%%\caption{\label{gammalist}List of detected $\gamma-$rays from different nuclei with half-life and $\gamma-$ray intensity}
%%\footnotesize
%%\begin{tabular}{@{}ccc}
%%\br
%%Produced nuclei&Half$-$life (T$_{1/2}$)~\cite{BROWNE2009507, BROWNE20102425}&Detected $\gamma-$rays \\&& with %%intensity~\cite{BROWNE2009507, BROWNE20102425}\\
%%\mr
%%$^{145}$Eu&5.93(4) days&893 (66\%)\\&&653.5(15\%)\\
%%\mr
%%$^{65}$Zn&243.93(9) days&1115.5(50\%) \\

%%\br
%%\end{tabular}\\
%%\end{table} 

A careful background subtraction was performed using CERN ROOT Data Analysis Framework~\cite{brun1997root} to find the peak counts of $\gamma-$ray of interest. The peak areas were determined by subtracting the average background estimated from the neighboring lower- and higher-energy regions. The measured cross sections carried an uncertainty of about 25\%. This was calculated by considering uncertainties from all possible sources like $\gamma-$intensity, decay constant, target thickness, standard source activity and also from the statistical and fitting of peak area and detector efficiency curve. Detailed of these uncertainties are discussed as a form of covariance matrix in a separate section.

\subsection{Statistical model estimation}
\begin{figure}[h]
\centering
\includegraphics[clip, trim=0.0cm 0.0cm 0.0cm 0.0cm,width=0.75\textwidth]{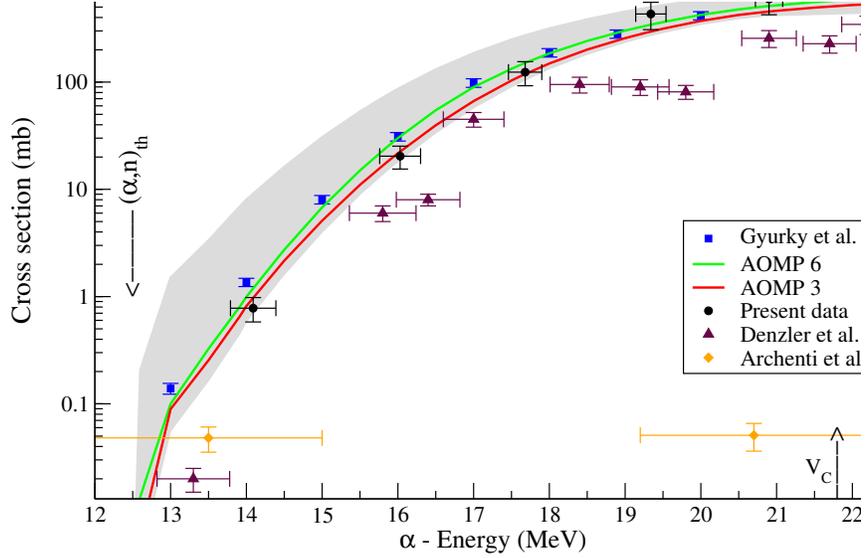}
\caption{The experimental cross-sections and Hauser-Feshbach calculations obtained from TALYS 2.0 are presented. Grey shaded area denotes the theoretically calculated cross section from TALYS by varying all OPM, NLD and GSF combinations. Additional experimental data are taken from literature~\cite{gyurky2023cross, Denzler1995,archenti1989alpha}.}
\label{cross_fig}
\end{figure}

The measured $^{144}$Sm($\alpha$,n)$^{147}$Gd reaction cross sections (Figure~\ref{cross_fig}) were compared with the theoretically obtained values from statistical Hauser-Feshbach model code TALYS 2.0~\cite{koning2023talys}. These calculations were performed using 8 different alpha optical model potential (AOMP), 6 different level density parameters (LD) and 9 different $\gamma-$ray strength functions (GSF). A Total of 432 combination of parameters were used to find the complete theoretically predicted band in TALYS 2.0 (grey band in Figure~\ref{cross_fig}) for cross sections between the reported $\alpha-$energies. All the models used during calculation are listed in Table~\ref{Model_list}. 
Boundary of this grey shade indicates the maximum and minimum values obtained from the TALYS calculation. During the calculation it was found that cross section values are majorly influenced by the choice of AOMP and different combination of LD and GSF to that AOMP model does not change the value drastically. Two representative calculated curves that reasonably reproduce the present experimental data are shown in Figure~\ref{cross_fig}, labelled AOMP3 and AOMP6, together with previously reported measurements~\cite{gyurky2023cross, Denzler1995,archenti1989alpha}. The AOMP3 set corresponds to calculations employing the $\alpha-$optical potential of Demetriou and Goriely~\cite{demetriou2002improved}, the Constant Temperature and Fermi Gas (CTM) level-density prescription~\cite{gilbert1965composite, ericson1960statistical}, and the Kopecky–Uhl generalized Lorentzian $\gamma-$ray strength function~\cite{kopecky1990test}. Alternatively, AOMP6 utilizes the $\alpha-$optical potential of Avrigeanu \textit{et al.}~\cite{avrigeanu2014further}, the Gogny–Hartree–Fock–Bogoliubov microscopic level densities from numerical tables~\cite{goriely2006microscopic}, and the Brink–Axel Lorentzian description of the $\gamma-$ray strength function~\cite{brink1957individual, axel1962electric}.\\

\begin{table}
\centering
\caption{\label{Model_list}Different models used during TALYS calculation. A total of 432 combinations from these models have been tried to find the cross section predictions.}
\footnotesize
\begin{tabular}{@{}lll}
\br
Parameter&Models&Details\\
\mr
Optical potential& AOPM 1& \parbox[t]{10cm}{Watanabe folding approach with Koning-Delaroche nucleon potentials~\cite{koning2003local, watanabe1958high}}\\
& AOPM 2& Alpha potential of McFadden and Satchler~\cite{mcfadden1966optical}\\
& AOPM 3& Alpha potential of Demetriou and Goriely~\cite{demetriou2002improved}, table 1 \\
& AOPM 4& Alpha potential of Demetriou and Goriely~\cite{demetriou2002improved}, table 2 \\
& AOPM 5&Alpha potential of Demetriou and Goriely~\cite{demetriou2002improved}, dispersive model\\
& AOPM 6& Alpha potential of Avrigeanu \textit{et al.}~\cite{avrigeanu2014further}\\
& AOPM 7& Alpha potential of Nolte \textit{et al.}~\cite{ nolte1987global}\\
& AOPM 8& Alpha potential of Avrigeanu \textit{et al.}~\cite{avrigeanu1994global}\\
\mr
Nuclear level density&LD 1& Constant temperature and Fermi gas model (CTM)~\cite{gilbert1965composite,ericson1960statistical}\\
&LD 2& back-shifted Fermi-gas model (BFM)~\cite{dilg1973level}\\
&LD 3& Generalised Superfluid model(GSM)~\cite{ignatyuk1979kk,ignatyuk1993density}\\
&LD 4& \parbox[t]{10cm}{Skyrme-Hartree-Fock-Bogoluybov level densities from numerical tables (microscopic model)~\cite{goriely2006microscopic}}\\
&LD 5& \parbox[t]{10cm}{Gogny-Hartree-Fock-Bogoluybov level densities from numerical tables (microscopic model)~\cite{goriely2006microscopic}}\\
&LD 6& \parbox[t]{10cm}{Temperature-dependent Gogny-Hartree-Fock-Bogoluybov level densities from numerical tables (microscopic model)~\cite{hilaire2012temperature}}\\
\mr
$\gamma-$ray strength function&GSF 1& Kopecky-Uhl generalized Lorentzian~\cite{kopecky1990test}\\
&GSF 2& Brink-Axel Lorentzian~\cite{brink1957individual,axel1962electric}\\
&GSF 3& Hartree-Fock BCS tables~\cite{goriely2002large}\\
&GSF 4& Hartree-Fock-Bogoliubov (HFB) tables~\cite{goriely2004microscopic}\\
&GSF 5& Goriely’s hybrid model~\cite{goriely1998radiative}\\
&GSF 6& Goriely T-dependent HFB~\cite{goriely2004microscopic}\\
&GSF 7& Temperature-dependent Relativistic Mean Field (RMF) model~\cite{daoutidis2012large}\\
&GSF 8& \parbox[t]{10cm}{Gogny-Hartree-Fock-Bogoliubov model with the quasiparticle random phase approximation (QRPA)~\cite{goriely2018gogny}}\\
&GSF 9& \parbox[t]{10cm}{Simplified Modified Lorentzian (SMLO) by Stephane Goriely and Vladimir Plujko~\cite{plujko2019description}}\\
\br
\end{tabular}
\end{table}

\subsection{Covariance analysis}
The reaction cross sections obtained in this study have uncertainties due to various factors like statistical uncertainty from counting, uncertainty in the target thickness, detector efficiency etc. All these contributes towards a cumulative uncertainty and these need to be considered cautiously. For this reason, a detailed covariance analysis is performed between all measured data points. 
A covariance matrix is a compact way of showing how uncertainties in different measured or fitted quantities are related to each other. Covariance matrix shown in Table~\ref{eff_fitTab12.5mm} and~\ref{eff_fitTab50mm} depicts the uncertainty in the fitting parameters as well as the correlation between parameters $\varepsilon_0$, $\varepsilon_1$ and $E_0$.
The cross section for a measurement \textit{i} is a function of a set of input quantities $x_1$, $x_2$, $x_3...x_\textit{k}$ (like peak count, efficiency etc.), 
\begin{equation}
\sigma_\textit{i} = \sigma_\textit{i}(x_1, x_2, x_3,....x_\textit{k}).
\label{cov1}
\end{equation}
The covariance between measured quantity and the input parameter is expressed as~\cite{smith2012experimental,lawriniang2019measurements,Bevington2003,otuka2017uncertainty,Cowan1998},
\begin{equation}
V_\sigma = \mathcal{J}~C~\mathcal{J}^T
\label{cov2}
\end{equation}
where $V_\sigma$ denotes the covariance matrix of the measured cross sections, $C$ represents the covariance of the input parameters, and $\mathcal{J}$ is the Jacobian matrix defined with elements $\frac{\partial\sigma_\textit{i}}{\partial x_\textit{k}}$.  Here, $\textit{i}$ indexes the number of measured cross sections, while $\textit{k}$ corresponds to the total number of input parameters or attributes appearing in Eq.~\ref{cross_eq3}.
The percentage uncertainties associated with the input parameters entering the cross-section expression are summarized in Table~\ref{rho_list}. The target thickness and counting statistics are considered uncorrelated, as independent targets were employed and the measurements were carried out individually for each cross section determination.

\begin{table}
\centering
\caption{\label{rho_list}Relative uncertainties of the input parameters entering the cross section formula}
\footnotesize
\begin{tabular}{@{}lc}
\br
Parameters&Uncertainty (\%)\\
\br
Beam current	& $\sim 5\%$ 	\\ \\

Detector Efficiency& \parbox[t]{3cm}{$2.82\%$ for 50 mm $3.56\%$ for 12.5 mm} 	\\ \\

$\gamma-$transition intensity& $\sim 4\%$ 	\\ \\

Half-life	& $\sim 0.32\%$ 	\\ \\

Target thickness	& $\sim 15-25\%$ 	\\ \\

Counting statistics	& $\sim 1\%$ 	\\ \\

\br
\end{tabular}\\
\end{table}

After construction of the complete covariance matrix with elements $V_{\textit{ij}}$, it was normalized to obtain the correlation matrix (\textbf{$\Gamma$}), whose elements are given by
\begin{equation}
\Gamma_{\textit{ij}}=\frac{V_{\textit{ij}}}{\sqrt{V_{\textit{ii}}~V_{\textit{jj}}}}.
\label{cov8}
\end{equation}
Table~\ref{xs_list} summarizes the measured cross sections together with the full covariance matrix and the associated correlation matrix

\afterpage{
\begin{landscape}
\begin{table}
\centering
\caption{\label{xs_list}Measured cross-sections with covariance matrix.}
\footnotesize
\begin{tabular}{@{}cc|lllll|lllll}
\br
\parbox[t]{2.5cm}{Energy in MeV (E$_{lab}\pm 1\sigma$)}&\parbox[t]{3cm}{Cross section in mb ($\sigma \pm \Delta \sigma$)} &&Covariance& matrix&&&&Correlation& matrix (\%)\\
\br
\vspace{0.5cm}
20.90	$\pm$	0.18	&	594.61	$\pm$	154.38	&	23834.392	&&&&			 &\parbox[t]{1.2cm}{100}					\\
\vspace{0.5cm}
19.34	$\pm$	0.20	&	431.44	$\pm$	112.03	&	1258.358	&	12550.575	&&&&	7.28	&	\parbox[t]{1.2cm}{100} \\ \vspace{0.5cm}
17.68	$\pm$	0.22	&	124.53	$\pm$	32.45	&	378.676		&	274.761		&	1052.974&&&	7.56	&	7.56	&	100\\
\vspace{0.5cm}
16.03	$\pm$	0.27	&	20.52	$\pm$	5.35	&	62.382		&	45.263		&	13.738		&	28.576	&&	7.56	&		7.56	&	7.92	&	100 \\ 
\vspace{0.5cm}
14.09	$\pm$	0.30	&	0.79	$\pm$	0.20	&	2.387		&	1.732		&	0.526		&	0.087	&	0.042	&	7.56	&	7.56	&	7.92	&	\parbox[t]{1.2cm}{7.92}	&100 \\ 
\br
\end{tabular}
\end{table}
\end{landscape}
}

\section{Discussion and Conclusion}
In this article, the reaction cross section of $^{144}$Sm($\alpha$,n)$^{147}$Gd has been measured at sub-Coulomb barrier energies ranging between 14 and 21 MeV. It was done using the stacked foil activation technique followed by the off-line gamma ray spectroscopy measurement. A detailed GEANT4 simulation has been carried out to determine the energy uncertainty due to the combined effect of initial beam spread from the cyclotron and straggling in the degrader and target foil including Al backing. This study presents, for the first time, an extensive uncertainty evaluation in the form of covariance and correlation matrices for this reaction. The obtained cross section values are then compared with the previously available data from literature and also with the theoretical predictions obtained from Hauser Feshbach statistical model code TALYS 2.0. In the calculations, the predicted cross sections were found to be more sensitive to the choice of the $\alpha-$optical model potential than to the selection of the level-density prescription or the $\gamma-$ray strength function. Among the used potentials, AOMP1 and AOMP2 reproduce the highest three data points to some extent but show a slight overestimation at lower energies. The potentials AOMP3 to AOMP6 offer a reasonable description of the lower-energy data points, although they tend to slightly underestimate the highest two data points. In contrast, AOMP7 and AOMP8 provide a better reproduction of the upper data points but significantly overestimate all lower-energy measurements. In the figure~\ref{cross_fig}, the results from all parameter combinations are displayed as a grey band, while the curves corresponding to the representative parameter sets AOMP3 and AOMP6 are shown separately for clarity. The covariance and correlation analysis indicates that the measured cross sections exhibit mutual correlations at the level of approximately 7$-$8\% among themselves (Table~\ref{xs_list}).
\section*{Acknowledgement}
The authors express their profound gratitude to Prof. Chandi Charan Dey for supplying the enriched target material, and to Prof. Chandana Bhattacharya for her invaluable contribution to the successful completion of the experiment. We extend our sincere thanks to Mr. Sudipta Barman and the other members of the workshop at the Saha Institute of Nuclear Physics, Kolkata for their support. We also acknowledge the assistance provided by the staff of the K130 Cyclotron, VECC Kolkata, and Mr. A. A. Mallick from the Analytical Chemistry Division, BARC–VECC during the irradiation process.

SS gratefully acknowledges the financial support received from the Council of Scientific and Industrial Research (CSIR), Government of India, through the Senior Research Fellowship (File No. 09/489(0119)/2019-EMR-I). 

\section*{Data availability statement}
The data that support the findings of this study are available upon reasonable request from the authors.
\section*{References}

%\bibliographystyle{iopams.sty}
%\nocite{apsrev41Control}
\bibliography{ref_file}

\end{document}